\documentclass[authoryear, 5p]{elsarticle}

\usepackage{moreverb,url}
\usepackage[colorlinks,bookmarksopen,bookmarksnumbered,citecolor=green!60!black,urlcolor=blue!80!black]{hyperref}
\usepackage{xcolor}

\usepackage{lineno}
\usepackage{setspace}
\modulolinenumbers[1]
\usepackage{listings}
\usepackage{amsmath}
\usepackage{booktabs}
\newcommand{\fastFADE}{DARF}
\begin{document}
\begin{frontmatter}
\title{DARF: A data-reduced FADE version for simulations of speech recognition thresholds with real hearing aids}
\author[1]{David H\"ulsmeier\corref{cor1}}
\ead{david.huelsmeier@uni-oldenburg.de}
\author[1]{Marc Ren\'e Sch\"adler}
\author[1]{Birger Kollmeier}

\address[1]{Medizinische Physik and Cluster of Excellence Hearing4all, CvO Universit\"at Oldenburg, 26129 Oldenburg, Germany}

\cortext[cor1]{Corresponding author}

\begin{abstract}
Developing and selecting hearing aids is a time consuming process which is simplified by using objective models.
Previously, the framework for auditory discrimination experiments (FADE) accurately simulated benefits of hearing aid algorithms with root mean squared prediction errors below 3\,dB.
One FADE simulation requires several hours of (un)processed signals, which is obstructive when the signals have to be recorded.
We propose and evaluate a \underline{da}ta-\underline{r}educed \underline{F}ADE version (\fastFADE{}) which facilitates simulations with signals that cannot be processed digitally, but that can only be recorded in real-time.
\fastFADE{} simulates one speech recognition threshold (SRT) with about 30 minutes of recorded and processed signals of the (German) matrix sentence test.
Benchmark experiments were carried out to compare \fastFADE{} and standard FADE exhibiting small differences for stationary maskers (1\,dB), but larger differences with strongly fluctuating maskers (5\,dB).
Hearing impairment and hearing aid algorithms seemed to reduce the differences.
Hearing aid benefits were simulated in terms of speech recognition with three pairs of real hearing aids in silence ($\geq$8\,dB), in stationary and fluctuating maskers in co-located (stat. 2\,dB; fluct. 6\,dB), and spatially separated speech and noise signals (stat. $\geq$8\,dB; fluct. 8\,dB).
The simulations were plausible in comparison to data from literature, but a comparison with empirical data is still open.
\fastFADE{} facilitates objective SRT simulations with real devices with unknown signal processing in real environments.
Yet, a validation of \fastFADE{} for devices with unknown signal processing is still pending since it was only tested with three similar devices.
Nonetheless, \fastFADE{} could be used for improving as well as for developing or model-based fitting of hearing aids.
\end{abstract}


\end{frontmatter}
\section{Introduction}
Hearing impairment imposes a barrier within the daily life of many people as it impedes conversations and the perception of (warning) sounds.
Hearing aids counteract loss of audibility and improve speech recognition.
However, finding the best hearing aid and fitting for an individual is difficult and time-consuming \citep{boymans2012audiologist, volker2018hearing}.
The development of hearing aids and algorithms is also time consuming and prone to errors when predicting their possible benefits for speech recognition with models, such as, e.g., the SII \citep{SII} or iSNR \citep{greenberg1993intelligibility} (e.g., \citet{Baumgartel2015} and \citet{volker2015comparing}, or see \citet{falk2015objective} and \citet{kollmeier2018functionality} for an overview).
These and more recent models that predict hearing device performance (e.g., HASPI \citealt{HASPI}, HASQI \citealt{HASQI}, STOI \citealt{STOI}) have three disadvantages:
First, they predict index values whose perceptual meaning is non-transparent, i.e., index values have to be mapped to human speech recognition performance in terms of either a percentage of correctly recognized words or a speech recognition threshold (SRT) which is the signal-to-noise ratio (SNR) with a 50\,\% recognition rate.
Second, most models often perform only well for specific tasks (e.g., only with stationary maskers), such that the task defines the model \citep{falk2015objective}.
Third, such models are intrusive \citep[by the definition of][]{schaedler2018objective}, i.e., they require some form of separable speech and noise signals.
This poses another obstacle when signal and noise need to be separated after processing which requires assumptions about the non-linear system \citep[see][for an overview]{schaedler2018objective}.

The process to develop, evaluate, and optimize hearing aids and their algorithms can profit from objective models that accurately predict benefits in a fast and convenient way.
The potential advantage of such simulations is even higher if models are used that predict the benefit of real devices.
Real devices typically show deviations from the idealized behavior of the respective underlying hearing aid algorithm.
That is, real signals are recorded and environment-specific modifications of the underlying combination of signal processing algorithms are performed.
These complex interactions between the acoustic environment, the acoustic paths, and the respective algorithm may even not be accessible when modeling hearing aids.
Therefore, several pitfalls are avoided by using real hearing devices.
Yet, at the time of writing, no model allows to directly simulate benefits of real hearing aids that can be worn and be directly used by hearing impaired listeners, at least in a practicable time frame.

To be useful for individual recommendations, the individual benefit of hearing aids has to be predicted accurately which imposes a difficult task:
Only a limited success in predicting hearing aid benefits was achieved on the individual prediction task for hearing aid algorithms \citep[e.g.,][]{Baumgartel2015, falk2015objective, schaedler2018objective}.
Further, the currently available estimates of average improvement prediction by index values \citep[e.g.,][]{kates2018using} are not yet validated for the individual aided performance prediction with real hearing devices.
Thus, the actual hearing aid benefit remains unknown.
Recent advances were reported by \citet{schaedler2018objective, schaedler2020individual} who used automatic speech recognition (ASR) systems to predict hearing aid algorithm performance, and by \citet{fontan2020improving} who used ASR systems to predict optimized hearing aid gain settings.
\citet{schaedler2018objective} accurately predicted the benefit from several binaural noise reduction algorithms for normal listeners and the average of hearing-impaired listeners.
\citet{fontan2020improving} qualitatively predicted the intelligibility for different hearing aid gain settings to prescribe an optimum setting, while \citet{schaedler2020individual} accurately predicted the benefit of virtual hearing aids implemented in the (open) Master Hearing Aid \citep[MHA,][]{grimm2006mha, herzke2017open}.
Yet, all of these approaches require access to the hearing aid algorithms which complicates their use with real devices as ``black box'', i.e., when the signals cannot be generated and processed offline, but rather have to be recorded in real time.

\citet{schaedler2020individual} used the framework for auditory discrimination experiments \citep[FADE,][]{schaedler2016simulation} to simulate human speech recognition to overcome limitations of traditional models \citep[available online, see][]{FADE_online}.
The approach does not require separated speech and noise signals and it was shown to accurately predict SRTs and algorithm benefits of listeners with normal and impaired hearing in a number of stationary and non-stationary noise and aided listening conditions \citep{schaedler2016simulation, schaedler2018objective, schaedler2020individual}.
That is, \citet[][see Tab. 3]{schaedler2020individual} predicted speech recognition benefits (SRTs measured in quiet, stationary, and fluctuating maskers) of three different hearing aid algorithms (amplification, compression amplification, and compression amplification with a noise suppressing beamformer) of eleven listeners (normal hearing to moderate hearing impairment) with prediction errors of less than 3\,dB (i.e., close to the test-retest accuracy).

The downside of the training procedure is that FADE requires several hours of mixed speech and noise signals.
This makes it difficult to use FADE with any algorithm or device that can only process signals in real time.
For example, simulation performed by \citet{schaedler2018objective} or \citet{schaedler2020individual} would allow for two to three simulations per day since they required about nine hours of mixed signals to simulate one SRT.
However, many of the recordings are discarded at some stage of the SRT estimation process while only signals mixed at one training SNR and two test SNRs, i.e., 50 minutes of signals, currently provide the simulation outcome.

The aim of this paper is to reduce the number of signals required for one simulation with FADE to facilitate simulations of hearing aid benefits of \emph{real} devices.
An objective, non-intrusive simulation method for aided speech recognition performance that does not require any reference empirical speech recognition measurement to make simulations while providing plausible simulation outcomes is desirable which operates on an amount of speech data which is as small as possible.
Yet, non of the aforementioned models fulfills all of these criteria.
Therefore, we propose a version of FADE which requires less data that facilitates simulations in which signals are required to be processed in real-time (denoted as data-reduced FADE, \fastFADE{}).
Further, \fastFADE{} is tested for the accurate simulation of aided and unaided speech recognition performance.
Note that \fastFADE{} is currently limited to simulations with matrix sentence tests \citep{kollmeier2015multilingual} contrary to FADE, which can be used to simulate outcomes of other speech tests and psychoacoustic tests \citep[][]{huelsmeier2018extension}.
Such an approach might reduce the time required to find hearing aids for individuals and accelerate the development of hearing aids in an objective and evidence-driven way.

The accuracy of the \fastFADE{} approach can be shown by comparing it with simulations of the original FADE model, as well as with data from current literature.
For that purpose, measured SRT data and FADE simulations with normal hearing \citep{schaedler2016simulation, schaedler2016microscopic}, hearing impairment \citep{huelsmeier2020chat, wardenga2015you}, and hearing impairment together with hearing aid algorithms \citep{schaedler2020individual} were used.
However, at the time of writing no appropriate empirical data (i.e., that of listeners with similar hearing loss using the same hearing aids with the same configuration) were available for comparison with simulations with real hearing aids.
Yet, when the plausibility of the model approach can be demonstrated, it---in theory---facilitates model-based hearing aid fitting or its application as an assistance during the hearing aid development process.

Here, aided and unaided speech recognition performance is simulated with the (German) matrix sentence test, which can be used to reliably measure SRTs \citep[test-retest reliability below 1\,dB,][available in more than 20 languages]{kollmeier2015multilingual}.
The median test-retest reliability below 1\,dB for the German matrix sentence test was determined by \citet{wagener2005sentence} across a group of ten listeners with normal hearing, and across a group of ten listeners with impaired hearing.
After accounting for a training effect, \citet{wagener2005sentence} measured the listeners' SRTs using an adaptive procedure \citep{brand2002efficient} in which 30 sentences were presented to the listeners.
Similar test-retest reliabilities were found for matrix sentence tests in other languages \citep[e.g.,][]{warzybok2015development, jansen2012comparison}

To examine if the aim of the paper was reached, i.e., if the reduction of the number of signals required for one simulation with FADE was successful to facilitate simulations with real hearing aids, the following research questions were tested:
\begin{itemize}
  \item Is \fastFADE{} suitable to accurately predict aided and unaided speech recognition performance in a reasonable time span?
  \item Does \fastFADE{} predict plausible benefits for real hearing aids?
\end{itemize}

\section{Methods}
\subsection{Model approach}
\subsubsection{Framework for Auditory Discrimination Experiments}
\begin{figure*}[htb]
\centering
\includegraphics[width=0.99\textwidth]{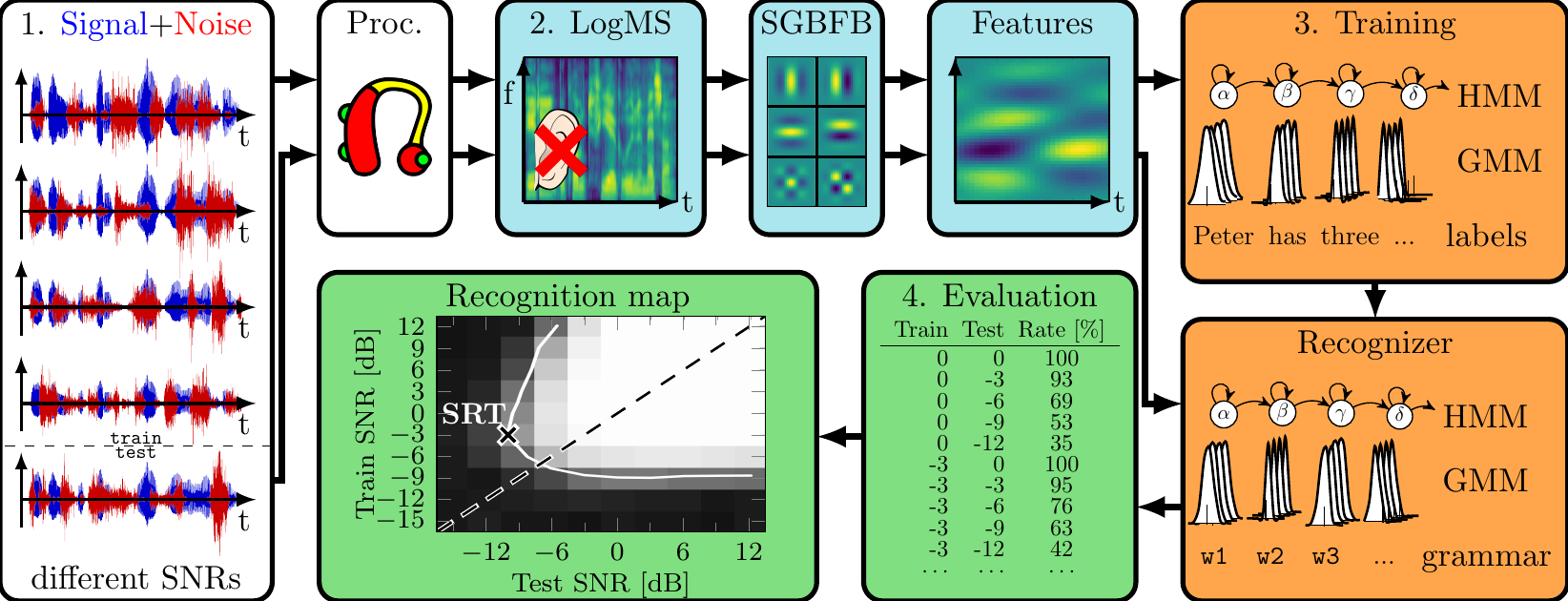}
\caption{FADE processing scheme to predict one SRT:
         1) Speech and noise signals are mixed at different SNRs with excerpts of the masker.
            Subsequently, the mixed signals can be processed with any algorithm or hearing device.
         2) Logarithmically scaled Mel spectrograms (LogMS) of the signals are calculated and separable Gabor filterbank features \citep[SGBFBs][]{schaedler2015separable} are extracted from them.
           Hearing impairment is applied when required (indicated by the crossed out ear).
         3) Speech recognizers are trained and tested for each SNR.
         4) The correct recognition rate for each training and test SNR pair is evaluated and depicted in a recognition map (0\,\% indicated by black, 100\,\% indicated by white).
         The lowest interpolated test SNR which yields a 50\,\%-correct rate is used as modeled SRT.%
         }
\label{fig:fade}
\end{figure*}

FADE is a model that simulates the outcome of auditory discrimination experiments, i.e., of psychoacoustic or of speech recognition experiments.
Its modeling scheme with regard to speech recognition experiments is depicted in Figure \ref{fig:fade}.
Any modification of the model enforces retraining it (e.g., using different speech signals, a different masker, a different signal processing, or other parameters for taking into account hearing impairment).
Hence, every SRT simulation conducted with FADE requires the execution of every step of the modeling scheme.

First, speech and noise signals are mixed at different SNRs for training and testing the hidden Markov model with Gaussian mixture model (GMM-HMM) speech recognizers.
FADE uses one recognition system per training SNR to simulate the outcome of speech recognition experiments based on the (German) matrix sentence test \citep{kollmeier2015multilingual}.
About 40 minutes of training signals (i.e., 960 noisy matrix sentences) and about five minutes of test signals (i.e., 120 noisy sentences) at each SNR are required with the standard parameters of FADE.
Thus, a standard simulation with assumed normal hearing and eleven SNRs requires about nine hours of train and test signals.
These signals may be processed digitally with any signal processing algorithm or they may be processed analogously by playing and recording the signals.
Next, spectro-temporal representations of the signals, i.e., logarithmically scaled Mel spectrograms (LogMS), are calculated for each signal.
These spectrograms can be modified to incorporate hearing impairment.
Then, features are extracted from the (modified) LogMS and used for training and testing the speech recognition systems of FADE \cite[typically separable Gabor filterbank features are used for that purpose][]{schaedler2015separable}.
Hence, features extracted from spectrograms with incorporated hearing impairment are used for training and testing.
At last, the recognition rate is calculated for each training and test SNR which is then depicted in a recognition map (see Fig. \ref{fig:fade}).
There, the recognition performance is plotted as a function of the SNR used for training different speech recognition systems against the test SNRs.
The SRT is then estimated as the lowest interpolated test SNR which yields a 50\,\% correct recognition score.

Hearing impairment was taken into account by modifying the LogMS.
That is, signal energy below the absolute threshold was removed from the spectral representation when hearing loss was used \citep[][]{kollmeier2016sentence, huelsmeier2020params}.
Therefore, the absolute threshold was converted to dB\,SPL using \citet{iso2003226} and a freefield to eardrum correction \citep{shaw1985transformation}.
When applicable, supra-threshold processing deficits were taken into account by adding a level uncertainty to the LogMS \citep[see][for details of the implementation]{huelsmeier2020params}.

Binaural hearing was taken into account by calculating and concatenating separate features when two signal channels are present, i.e., one for the left and another for the right ear \citep[][]{schaedler2018objective, schaedler2020kain}.
These features were used to train one GMM-HMM ASR model per SNR, i.e., similar to the modeling procedure when using signals with one channel.
With this approach, the recognition systems automatically use the optimal statistics to recognize words, but without taking into account any binaural interactions \citep[comparable with better ear listening][]{hauth2018modeling}.
Note that FADE's simulations do not explicitly take into account the overall presentation level (closely related to loudness) or listening effort.
For more details on FADE and its implementations, please refer to \citet{schaedler2016simulation}, \citet{schaedler2018objective} and/or \citet{schaedler2020individual}.

Here, aided and unaided speech recognition performance is simulated with the matrix sentence test, which can be used to reliably measure SRTs \citep[test-retest reliability of 1\,dB,][available in more than 20 languages]{kollmeier2015multilingual}.
Matrix sentence tests are typically used in clinical practice \citep{hoth2017aktuelle, bundesausschuss2012hilfsmittel}, but
have also been successfully used in many studies for assessing the benefit for a large range of hearing devices (e.g., hearing aids \citealt{neher2017speech}, assistive listening devices \citealt{rennies2017extension, ihler2016prediction} or cochlea implants \citealt{williges2015spatial}).
These tests provide syntactical fixed and semantically unpredictable sentences which are composed of a name, verb, number, adjective and an object, e.g., ``Peter has five wet chairs'', where each word class (e.g., name) has ten alternatives (50 words in total).
Hence, these tests use simple grammar structures in combination with limited semantic and linguistic complexity, such that little cognitive capabilities are required to perform them.
Thus, mainly the ability to recognize words that are embedded in a natural-sounding sentence is examined with this test.

Further, matrix sentence tests are an ideal test for machine-learning-based objective prediction methods \citep{schaedler2016simulation}.
That is, their structure and vocabulary size simplify the ASR system's language model, and their acoustic model is not required to generalize across different speakers, genders, dialects, and other speaker-dependent idiosyncrasies.
Therefore, simulated SRTs of the (German) matrix sentence test are used as the prime measure to quantify speech recognition performance throughout this study.
Note, that learning the idiosyncrasies of a speech test leads to highly specialized ASR-systems that cannot generalize across, e.g., speakers, genders, or dialects.
Yet, listeners accustomed to the speech test can also learn such idiosyncrasies, which may be an explanation of the training effect observed for Matrix sentence tests \citep{kollmeier2015multilingual}.
Therefore, FADE simulates the performance of a listener highly accustomed to the given speech test and masker.
\subsubsection{Modified FADE (\fastFADE{})}\label{sec:fastFADE}
The standard FADE model is comparable to a brute-force method that uses a broad range of SNRs and a large amount of signals to estimate and assess one single SRT.
To reduce the required amount of signals for one simulation, the standard FADE version was modified based on assumptions about the model.
These modifications are i) model initialization, ii) SRT approximation, iii) SRT simulation, and iv) model evaluation.
\begin{description}
 \item[i) Initialization]
   A rough initial estimate of the SRT is made based on the average hearing loss for frequencies below 1\,kHz and the noise level.
   This assumption was found to be representative for the German matrix sentence test and various degrees of hearing loss \citep{huelsmeier2020chat}.
   The estimated SRT is the maximum of either the average hearing loss for frequencies below 1\,kHz in dB\,SPL, the SRT dependent on the noise level in dB\,SPL (about -8\,dB\,SNR), or the SRT in silence (about 15\,dB\,SPL).
   The motivation for this step is to yield a SNR close to a 50\,\% recognition rate for different degrees of hearing loss and noise levels, such that less signals are required in the following to find the lowest SRT.
   The hard coded estimation is motivated by SRTs in silence and in stationary, unmodulated noise, and their relation to the average hearing loss \citep{plomp1978auditory, wardenga2015you, kollmeier2016sentence, huelsmeier2020chat}.

   Example: The noise is presented at 65\,dB\,SPL to a listener who has an average hearing loss of 60\,dB\,SPL across 250, 500, and 1000\,Hz.
   Therefore, the initial estimate of the SNR is at -5\,dB\,SNR since the listening test is not conducted in silence, but the average hearing loss exceeds the SRT in noise (-8\,dB\,SNR or 57\,dB\,SPL).
 \item[ii) SRT approximation]
   A pre-simulation is conducted where the SRT is approximated starting from the rough initial SRT estimate with matched-SNR training, i.e., when train and test SNRs  are equal (leading to the SRT estimate marked with a cross in panel \textit{a} of Fig. \ref{fig:fast-fade-1}).
   For this purpose, the recognition performance for a single word class of the matrix sentence test (here: names) is determined to reduce the training data by a factor of five while adjusting the SNR.
   The number of training signals is initially reduced by a factor of 8 from 960 to 120 mixtures.
   In total, this reduces the amount of training signals form 40 minutes to 75\,s per SNR.
   The number of test signals is initially reduced by a factor of 4.8 from 120 to 25 names, i.e., to 15\,s.
   Both train and test signals are used three times to obtain sufficient signals for training and to yield a resolution of 1\,\% (100 words) for the recognition process.
   The estimated SRT is then either interpolated between two SNRs if two rates between 25\,\% and 75\,\%-correct are found, or it is set to the SNR which yields a correct rate within (50$\pm$15)\,\%.
   In total, this step requires about 90\,s of signals for a simulation with one SNR and converges after 3.0$\pm$1.6 iterations (i.e., 4.3$\pm$2.3\,min).

   Example (cont.): The recognition rate is determined for matching training and test SNRs of -5\,dB\,SNR (initial SRT estimate).
   There, 32\% of the names are correctly recognized (i.e., 32 of 100 names).
   Since the rate was below 50\,\%, the SNR is adapted to 0\,dB which yields a recognition rate of 62\,\%.
   This concludes the approximation, resulting in an SRT of -3\,dB\,SNR since an interpolation between the two SNRs with recognition rates between 25\,\% and 75\,\% was possible (cross in panel a of Fig. \ref{fig:fast-fade-1}).
   \begin{figure}[htb]
   \centering
   \includegraphics[]{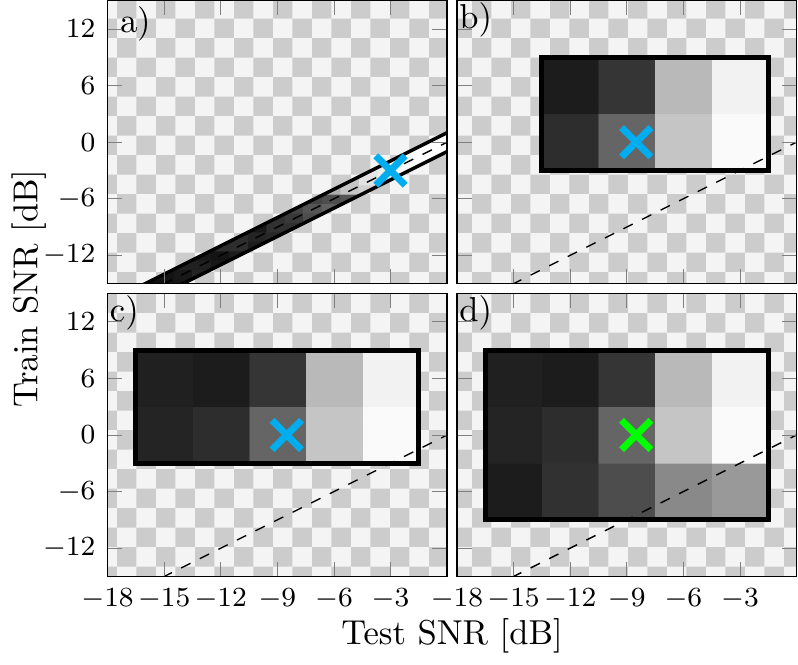}
      \caption{Recognition map during the adaptive search for the predicted SRT with \fastFADE{}:
               a) An SRT is estimated (cross) based on the recognition performance for one of the word classes of the matrix sentence test using matched SNR training.
               b) The recognition performance for a predefined range of training and test SNRs placed around the SRT estimated in the previous step is calculated.
               c) The training and test SNR range is adjusted until the lowest SNR with a 50\,\% recognition rate is found.
               d) The simulation converges (green cross).\\
               Note that the background checkerboard pattern indicates that no simulations were made for the SNR conditions.
               Hence, no information is available for these combinations of training and test SNRs.}
   \label{fig:fast-fade-1}
   \end{figure}
 \item[iii) SRT simulation]
   An adaptive search for the lowest SRT is started (panels \textit{b} to \textit{c} in Fig. \ref{fig:fast-fade-1}) which initially uses 120 train and 20 test sentences and thereby reduces the required signals per SNR from 45 to six minutes (training: 5\,min, test: 1\,min, factor 7.5):
   Two train and four test SNRs (10 + 4 min) are selected in dependence on the pre-simulation (panel \textit{a}), where the training SNRs are 3 and 9\,dB greater and the test SNRs are \mbox{-9}, \mbox{-6}, \mbox{-3}, and 0\,dB lower than the estimated SRT (panel \textit{b}).
   This SNR area is selected based on the typical pattern shown in all recognition maps:
   The lowest SNR with a 50\,\% recognition rate, i.e., the desired outcome of the simulation, has a greater or equal training SNR and a lower test SNR than found for the SNR on the diagonal of the recognition map that yields a 50\,\% recognition rate (e.g., as depicted in the block ``Recognition map`` in Fig. \ref{fig:fade}).
   Subsequently, the SNR area is adapted until a minimum SRT is found:
   \begin{itemize}
     \item Increase the test  SNRs by 3 dB if all recognition rates are lower than the target rate, or
     \item Decrease the test  SNRs by 3 dB if all recognition rates are greater than the target rate, or
     \item Decrease the test  SNRs by 3 dB if only one test SNR is below the SRT (e.g., panel b), and/or
     \item Increase the training SNRs by 6 dB if the recognition rate is found for the highest examined training SNR, or
     \item Decrease the training SNRs by 6 dB if the recognition rate is found for the lowest examined training SNR (e.g., panel c),
     \item Otherwise stop the simulation (e.g., panel d).
   \end{itemize}
   During these steps, the signals recorded at different SNRs are reused, i.e., each new test SNR is tested against all available and learned models for the training SNRs, and vice versa.
   Thereby, the signals are recorded while the other simulation steps (feature extraction, training and testing of the speech recognizers, and evaluation) run simultaneously whenever it is feasible, such that signals are recorded while running the remainder of the simulation process.
   Note that the simulation process alone requires about four minutes to finish.

   Example (cont.): Training signals are recorded at 0 and 6\,dB\,SNR (3 and 9\,dB grater than the approximated SRT of -3\,dB\,SNR), while test signals are recorded at -3, -6, -9, and -12\,dB\,SNR (panel b of Fig. \ref{fig:fast-fade-1}).
   An SRT was found at about -9\,dB\,SNR for the lowest recorded training SNR, while only one SNR below -9\,dB\,SNR was recorded for testing.
   Therefore, training signals are recorded at -6\,dB\,SNR and test signals are recorded at -15\,dB\,SNR.
   This concludes the simulation resulting in an SRT of about -9\,dB\,SNR since neither a lower training nor test SNR yielded a lower SRT.
 \item[iv) Evaluation]
   The amount of train sentences to determine the SRT is artificially doubled by applying multicondition training which generally improves the recognition performance \citep{hirsch2000aurora}.
   Therefore, signals recorded at two SNRs, e.g., at -3 and +3\,dB\,SNR, are used for training one combined speech recognizer.
   Note that this does not require to make any further recordings, but allows for more accurate predictions of conditions that normally require more data, e.g., simulations with fluctuating maskers.
   The simulated SRT is defined as the lowest SNR with a 50\,\%-correct recognition rate found with multicondition training.

   Example (cont.): The training SNRs of -6 and 0 dB\,SNR, and of 0 and 6 dB\,SNR are combined to two multicondition training conditions which are then tested with the recorded test SNRs in between -15 and -3\,dB\,SNR.
   This concludes the \fastFADE{} simulation yielding an SRT of about -8.5\,dB\,SNR.
\end{description}
\citet{schaedler2016simulation} found that 120 training sentences were sufficient to simulate speech recognition in stationary noise but recommended using 960 as this resulted in lower tone detection thresholds (note that FADE can simulate outcomes of psychoacoustic experiments contrary to \fastFADE{}).
Thus, SRTs simulated with \fastFADE{} and fewer signals can be expected to be similar to the SRTs simulated with FADE for simulations conducted with maskers that have little variance, or when the dynamic range of the signals is reduced to elevated absolute thresholds.
The differences between FADE and \fastFADE{} likely increase with fluctuating maskers, since using fewer train sentences results in a poorer representation of maskers with large variance by FADE's GMMs.
It is also less likely to find the best training SNR with \fastFADE{} since the training SNRs are sampled in 6\,dB steps (3\,dB steps with FADE).
Further, it can be expected that the amount of training signals affects the total amount of signals required for one simulation more than the amount of test signals.
\subsection{Experiment 1: Benchmarks with digital simulation}
In order to select suitable conditions to compare the simulation results of FADE and \fastFADE{} and to estimate the minimum amount of training and test data to be used by \fastFADE{}, the following requirements had to be fulfilled:
First, the number of train and test signals should provide plausible simulation outcomes with various maskers under the constraint that the simulations do not require much more time than a visit to an audiologist, i.e., about 30 minutes of audio material per simulation.
This facilitates to run one simulation with the listener's own hearing device during the appointment.
Second, the baseline for simulations, i.e., SRT simulations of listeners with normal (and impaired) hearing without any hearing aid, should be accurate (i.e., the outcome of repeated simulations should be within the test-retest accuracy), and yield similar outcomes with both FADE versions.
Hence, those preliminary versions of \fastFADE{} were excluded from further considerations that did not produce simulations with a sufficiently small prediction error due to a too much reduced dataset.
Third, for the same reason, both FADE versions should predict similar aided hearing performance.
Note, all of the benchmark simulations were conducted digitally, i.e., no signal was required to be recorded for the simulations.

These benchmarks were selected to see how \fastFADE{} performs in comparison to FADE, i.e., to examine what can and cannot be done with sufficient accuracy with the data-reduced version.
The underlying hypotheses of all benchmarks is that \fastFADE{} performs less accurate and reliable than FADE when the task includes more random processes not attributable to speech, e.g., when fluctuating maskers are used in combination with normal absolute thresholds.
Otherwise, i.e., when the dynamic of the signal is reduced due hearing loss or more stationary maskers, the differences between FADE and \fastFADE{} should reduce.
This hypotheses is in line with \citet[][Fig. 9]{schaedler2016simulation} who found that using 12 samples per word for training instead of 96 (which is used in the standard FADE approach) still results in accurate SRT predictions, at least when a stationary masker is used.
\subsubsection{Benchmark 1: Number of training and test signals}
The effect on the SRT of the number of train sentences used with \fastFADE{} was examined for multiples of 120 sentences (120, 240, ...,960), which equals one to eight times the complete sentences of the German matrix test.
The number of test sentences was examined for multiples of 20 sentences (20, 40, ..., 120), which equals one to six test lists used for measuring SRTs of human listeners.
The prediction performance with a given number of sentences was examined with a normal-hearing configuration and a fluctuating masker at 65\,dB\,SPL.
The masker had spectro-temporal gaps of up to 250\,ms \citep[icra5-250m,][]{Wagener2006}, and presents a condition that is difficult to simulate.
Listeners with normal hearing typically reach SRTs of less than -19.0\,dB\,SNR \citep[][]{hochmuth2015influence, wagener2005sentence} in this condition.
Simulations with other maskers (no masker, stationary icra1m, multitalker babble) were also conducted and can be found in the supplemental material.
Each combination of the number of train and test sentences was simulated 64-fold.
The average SRT difference to the standard FADE approach as well as the average simulation time were reported.

Due to the \fastFADE{} approach, varying numbers of training and test SNRs were used in each simulation.
However, the average time required for one simulation indicates how many SNRs were used for training/testing.
For example, playing and recording 20 sentences takes about one minute.
Therefore, a simulation that requires about 25 minutes of audio signals and which was conducted with 120 training sentences (five minutes per SNR) and 20 test sentences (one minute per SNR) uses---on average---three SNRs for the SRT approximation (4.5 minutes, step ii of Sec. \ref{sec:fastFADE}) and three training SNRs (15 minutes) in combination with five test SNRs (5 minutes) during the SRT simulation stage (step iii of Sec. \ref{sec:fastFADE}.

Optimum parameter settings were determined by finding an optimum accuracy speed tradeoff (AST, arbitrary units).
The AST was calculated as the multiplication of the simulated SRT's standard deviation ($s$), bias to simulations with standard FADE ($b$), and simulation time ($t$), and is normalized with their units (1\,dB$^{2}\cdot$1\,s):
\begin{eqnarray}
    \label{eq:AST}
    AST = \frac{\text{max}(1 \text{dB},s) \cdot \text{max}(1 \text{dB},b) \cdot t}{1 \text{dB}^{2}\cdot 1 \text{s}}
\end{eqnarray}
The lower limits of the simulated SRT's standard deviation and of the bias, i.e., the accuracy, were set to 1\,dB, which is close to the measurement accuracy of matrix sentence tests.
Thus, low ASTs can be found when \fastFADE{} and FADE yield the same outcome, the standard deviation of \fastFADE{}'s simulations was low, and/or when the simulations converged fast.
\subsubsection{Benchmark 2: Unaided hearing}
\fastFADE{} was compared to FADE by using simulated SRTs in different stationary \citep[icra1f, icra1m, and test specific noise, taken from][]{schaedler2016simulation, schaedler2016microscopic}, multitalker \citep[babble, taken from][]{schaedler2016microscopic}, and fluctuating maskers \citep[icra4-250m, and icra5-250m, taken from][]{schaedler2016microscopic}.
These baseline conditions were extended with one additional masker (multitalker: cafeteria) and silence.
Further, \fastFADE{} was compared with two empirical SRTs for the stationary icra1m and the fluctuating icra5-250m masker \citep[][average of ten listeners with normal hearing]{hochmuth2015influence}.
The masker levels were set to 65\,dB\,SPL.
Each of the simulations was conducted 512-fold.

The effect of hearing loss on the simulations was examined with SRTs of 315 ears with hearing impairment from \citet{wardenga2015you} and \citet{huelsmeier2020chat}.
The SRTs were measured monaurally using the stationary, test specific noise (tsn) at 65\,dB\,SPL.
The simulations were statistically examined with the root-mean-squared error \citep[RMSE,][]{armstrong1992error}, the bias (i.e., offset from the diagonal indicating a perfect correlation), and the coefficient of determination R$^{2}$ (i.e., the squared Pearson correlation coefficient).
\subsubsection{Benchmark 3: Aided hearing}
\fastFADE{} was compared to FADE by using the simulated aided SRTs of \citet{schaedler2020individual}, who measured and simulated SRTs of 18 listeners with different degrees of hearing loss.
They used the German matrix sentence test in silence and in stationary (icra1m) and fluctuating (icra5-250m) noise at 65\,dB\,SPL.
\citet{schaedler2020individual} examined different individualization strategies for hearing impairment.
Here, only the simulated data with an individualization for absolute and supra-threshold hearing impairment \citep[''AD`` by][]{schaedler2020individual} were considered, since this individualization gave the highest prediction accuracy.

For the simulation process, the noisy speech signals were digitally processed with the hearing aid algorithms used by \citet{schaedler2020individual} (including linear amplification, compression amplification, and compression amplification with a noise-suppressing beamformer) via the MHA before the LogMS were extracted and hearing impairment was applied.
This approach is equal to the approach of \citet{schaedler2020individual}.
The simulation accuracy was statistically examined with RMSE, bias, and the coefficient of determination R$^{2}$.
\subsection{Experiment 2: Benefit with real hearing aids}
\fastFADE{} was used to simulate SRTs with and without the provision of three \textit{real} hearing aid pairs from three leading hearing aid manufacturers.
The aim of this experiment was to examine the plausibility of the \fastFADE{} model's simulation outcome in combination with real devices.
Due to the simulation accuracy achieved by \citet{schaedler2020individual}, we expected that the simulated benefits were plausible.

For this experiment, hearing aids were put on a dummy head which was placed inside an anechoic chamber (Fig. \ref{fig:dummy}).
Speech and noise signals were presented via loudspeakers with a fixed noise level of 65\,dB\,SPL calibrated at the center position of the dummy head with an omnidirectional microphone.
SRTs were simulated with \fastFADE{} which steered the sound presentation and the recording.
A comparison with standard FADE was not feasible in a reasonable time frame, i.e., each simulation with FADE would require more than 9\,hours of signals (i.e., about one month of simulation time for all conditions).

\begin{figure}[htb]
   \centering
   \includegraphics[]{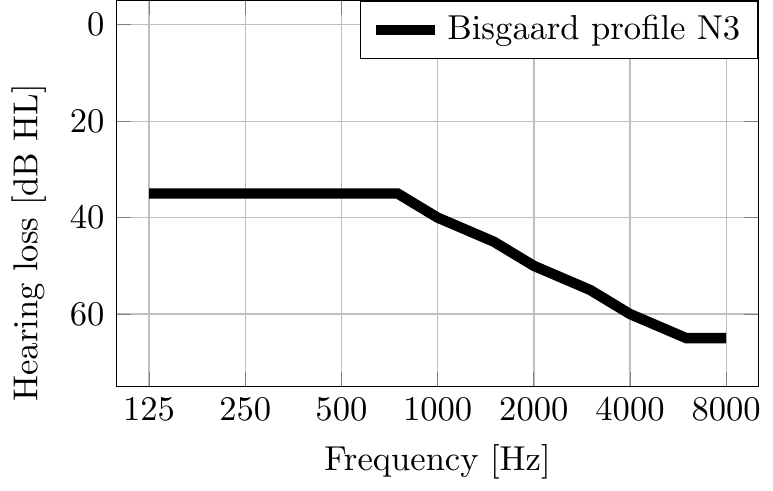}
      \caption{Hearing loss profile N3 (moderate high-frequency hearing loss) defined by \citet{bisgaard2010standard}.}
   \label{fig:bisgaard-n3}
   \end{figure}

\begin{figure}
 \centering
 \includegraphics[width=75mm]{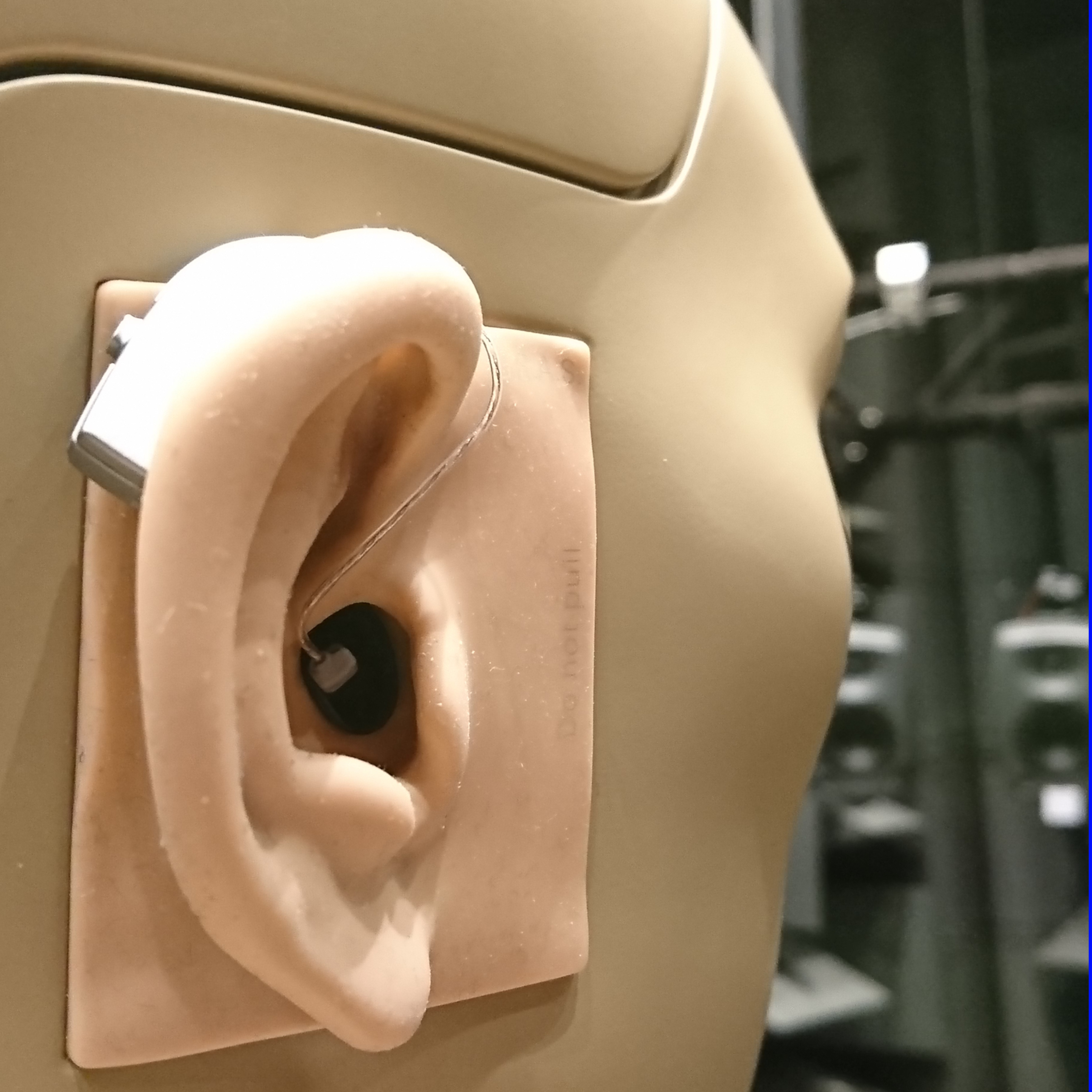}
 \caption{Right, aided ear of the dummy head.}
 \label{fig:dummy}
\end{figure}

The dummy head had two low noise microphones (left and right ear) with a sensitivity lower than the absolute threshold of normal hearing listeners \citep{wille2016iec}, with the exception of frequencies between 2.5 and 5.0\,kHz where it was about 2\,dB above the normal-hearing threshold.
The absolute hearing threshold on both ears was set to the profile N3 described by \citet{bisgaard2010standard}, i.e., a moderate high-frequency hearing loss which can be assumed to impede daily communication (Fig. \ref{fig:bisgaard-n3}).
The correct application of the absolute threshold was tested by recording pure tones and by analyzing the LogMS of the signals.
This procedure is similar to standard pure tone audiometry \citep{british2011recommended, american2005guidelines}, i.e., the level of pure tones were adjusted until the Mel spectra of the signals were not distinguishable from recordings without pure tones.
Binaural hearing was incorporated in FADE by concatenating the features extracted from the left and right signal channels.
This approach resembles (automatic) better ear listening \citep{schaedler2018objective, schaedler2020kain}, i.e., the system had to select which channel provides better recognitions.

The three (behind-the-ear) hearing aid pairs were on an approximately similar technological level (Tab. \ref{tab:hearing-aids}), i.e., their peak levels, maximum amplification, frequency range, microphone noise, and number of adjustable channels was alike.
Further, all devices had directional beamformers, noise suppression, and used double domes as ear-pieces.
The implementation of the hearing aid processing was not specified for any of the devices.
Therefore, the hearing aids can be thought of as black boxes that record, (non-linearly) process, and play sounds.
A simulation with normal-hearing thresholds was conducted to indicate if the hearing aids can restore simulated normal-hearing speech recognition performance.
Therefore, the normal-hearing condition can be considered as fourth ``hearing aid''.
\begin{table*}[htb]
 \small\sf\centering
 \caption{%
    Parameters of the commercial hearing aids employed.
    All devices used double domes as ear-pieces and applied beamformers and noise suppression algorithms.
    }
 \label{tab:hearing-aids}
 \begin{tabular}{c *{5}{c}}
   \toprule
          & peak level & max amplification & freq. range           & mic. noise & \# adjustable. \\
   Device & [dB\,SPL]  & [dB]     & [kHz]                 & [dB\,SPL]  & freq. channels \\
   \midrule
   A      & 126        & 61       & 0.1 - \phantom{1}9.5  & 20         & 16             \\ %
   B      & 129        & 70       & 0.1 - 10.0            & 23         & 20             \\ %
   C      & 124        & 69       & 0.1 - 10.0            & 21         & 15             \\ %
   \bottomrule
 \end{tabular}
\end{table*}

For each simulation, hearing aid pairs of one manufacturer and with the same fitting were placed on the dummy head.
The insertion of the hearing aid ear-pieces was controlled with a broadband noise to check if the same frequency response was found for the left and right ear.
The orientation of the dummy head, i.e., if the dummy head faced the front, was verified with an impulse played back from the loudspeaker in front of the dummy head and recording the time signal at both ears.
Thus, the dummy head was considered as facing to the front if the impulse reached the dummy head's ears at the same time.

An overview of the measurement conditions is listed in Table \ref{tab:measurment-conditions}.
The simulations were conducted in silence and in stationary (icra1m) and fluctuating (icra5-250m) maskers and with three spatial configurations.
That is, speech from the front (S$_{0}$) in combination with no masker, co-located maskers (S$_{0}$N$_{0}$), or the maskers being separated by 90$^{\circ}$ to the left (S$_{0}$N$_{90}$).
The spatial configurations were selected to examine if \fastFADE{} provides different benefits in binaural conditions which is one treatment goal of the German guidelines on (therapeutic) appliances \citep{bundesausschuss2012hilfsmittel}.
The hearing aid pairs were fitted with the NAL-NL2 fitting method \citep{keidser2011nal} and the methods proposed by the different manufacturers for each of the simulations.
The NAL-NL2 fitting method is based on the individual audiograms and tries to restore a comfortable loudness perception while improving speech intelligibility.
The manufacturers' fitting rules were based on the individual audiograms.
The fitting method of the manufacturer of hearing aid C aimed at preserving the natural sound of the hearing aid user's voice.
Otherwise, no further information about the manufacturers' fitting rules were provided.

The unaided SRT with hearing impairment was simulated eight-fold to provide a reference with low variance for determining hearing aid benefits.
Otherwise, and due to a lack of simulation time, standard deviations of the simulations were taken from the second benchmark experiment.
The standard deviation of the hearing aid benefits were calculated using Gaussian error propagation \citep{young1962experimental} and the standard deviations of the second benchmark experiment.
\begin{table*}[htb]
 \small\sf\centering
 \caption{Simulation conditions.
          Non-competing conditions are divided by horizontal lines, e.g., only one fitting can be used at a time, but any of the maskers can be used with it. 
          This results in 20 simulations with each hearing aid, i.e., about 12 hours of simulations with \fastFADE{} per hearing aid.
          N3: Hearing profile according to \citet{bisgaard2010standard}; NH: Normal hearing thresholds.
 }
 \begin{tabular}{l *{3}{c} }
  \toprule
                             & HI aided  & HI unaided & NH unaided  \\
  \midrule
  S$_{0}$                    & silence       & silence        & silence         \\
  S$_{0}$ N$_{0}$            & icra1m \& icra5-250m & icra1m \& icra5-250m  & icra1m \& icra5-250m   \\
  S$_{0}$ N$_{90}$           & icra1m \& icra5-250m & icra1m \& icra5-250m  & icra1m \& icra5-250m   \\
  \midrule
  Impairment                 &N3         &   N3       & NH          \\
  \midrule
  NAL-NL2 fitting            & x         &    -       & -           \\
  Manufacturer fitting       & x         &    -       & -           \\
  \midrule
  Num. of repetitions        & 2         &    8       & 2           \\
  \midrule
  \midrule
  Total num. of simulations  & 20        &   40       & 10          \\
  \bottomrule
 \end{tabular}
 \label{tab:measurment-conditions}
\end{table*}

\section{Results}
\subsection{Experiment 1: Benchmarks with digital simulation}
\subsubsection{Benchmark  1: Number of training and test signals}
\begin{figure}[htb]
\centering
 \includegraphics[]{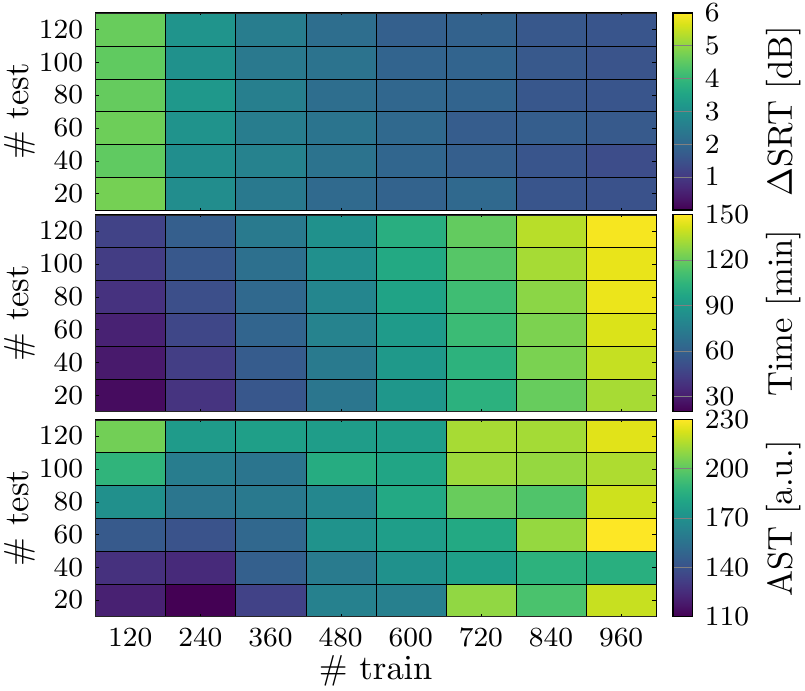}
 \caption{%
    Model performance comparison for the fluctuating icra5-250m masker for different numbers (\#) of train and test sentences (average of 64 simulations).
    Top: Differences in SRT between \fastFADE{} and FADE.
    Middle: Accumulated duration of all used sentences.
    Bottom: Accuracy speed tradeoff measure (AST, Eq. \ref{eq:AST}).
    }
 \label{fig:map-i5-ex}
\end{figure}

The differences between \fastFADE{} and FADE for the fluctuating icra5-250m masker in dependence on the number of training and test signals are depicted in Figure \ref{fig:map-i5-ex}.
As expected, the \fastFADE{} simulations generally showed 2 to 5\,dB higher SRTs.
The differences were mostly determined by the number of train sentences while changing the number of test sentences had a smaller effect on them.
The SRT differences were about 5\,dB when 120 training sentences were used and about 3\,dB with 240 training sentences.
The differences remained between 3 and 2\,dB when more training (or test) sentences were used.
Interestingly, a difference of about 1.5\,dB was found even when the standard number of train and test sentence (i.e., 120 test and 960 train sentences) was used (discussed later).

The time required for one simulation increased with the number of sentences, which was more affected by the amount of training sentences than by the amount of test sentences.
To assess the tradeoff between simulation accuracy and time for the simulation, the Accuracy Speed Tradeoff (AST, Eq. \ref{eq:AST}) was computed which is displayed in the lowest panel of Figure \ref{fig:map-i5-ex}.
The lowest AST was found for 240 training and 20 test sentences.
Further low ASTs were found when 120 or 240 training sentences were used in combination with 20 or 40 test sentences.

Even though employing 120 training sentences appears promising due to the least time consumption and a reasonable low AST, this condition provided the highest difference in SRT which may limit its usage for practical simulations.
However, these differences are expected to be much smaller in continuous noise than in the fluctuating noise employed here.
Therefore, further simulations were conducted with 120 and 240 training and 20 test sentences per SNR.
\subsubsection{Benchmark  2: Unaided hearing}
\begin{figure*}[htb]
\centering
 \includegraphics[]{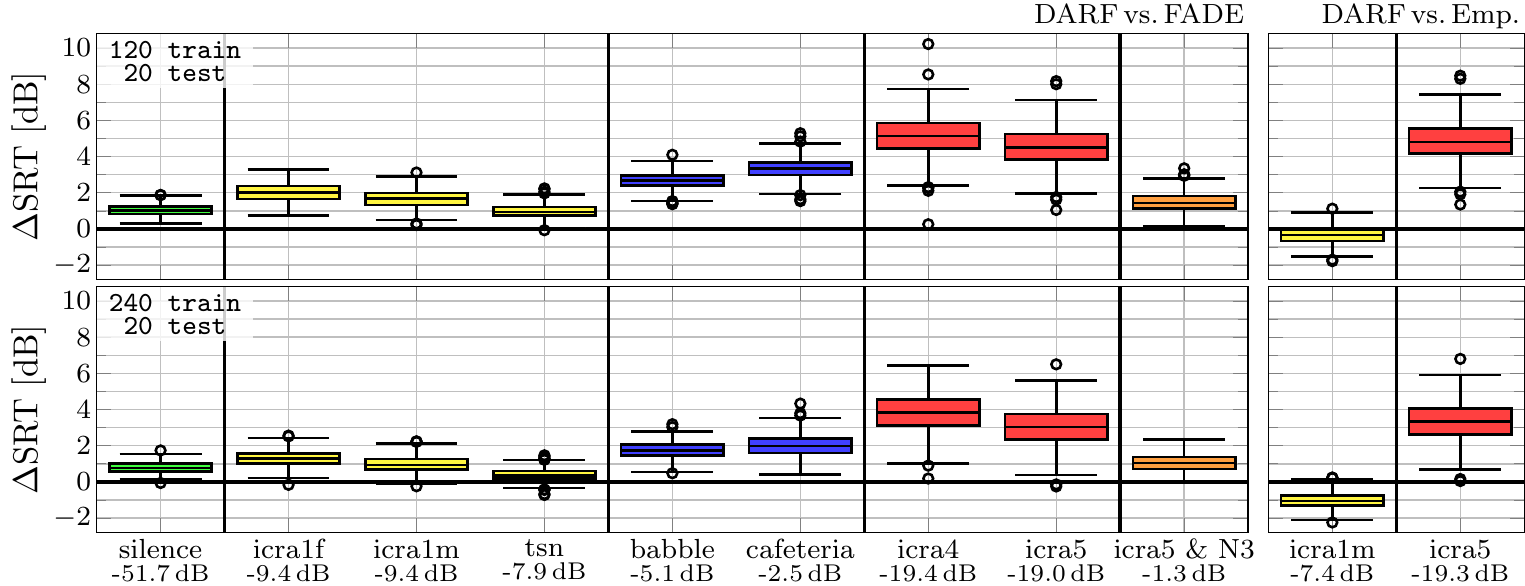}
 \caption{%
    Differences between SRTs ($\Delta$SRT) simulated by \fastFADE{} and FADE (left panels, \fastFADE\,vs.\,FADE), and differences between SRTs simulated by \fastFADE{} and two empirical SRTs measured by \citet{hochmuth2015influence} (right panels, \fastFADE\,vs.\,Emp.) based on 512 \fastFADE{} simulations for each masker.
    The numbers below the masker abbreviations denote the simulated SRTs with FADE or the empirical SRTs relative to 65\,dB\,SPL (i.e., the noise level).
    The colors indicate from left to right no (green: silence), stationary (yellow), multitalker (blue), or fluctuating (red) maskers.
    Orange indicates simulations with the fluctuating icra5-250m maskers and a moderate hearing loss \citep[N3,][]{bisgaard2010standard}.\\
    The boxplots show the median and the upper and lower quartiles (i.e., the 25\textsuperscript{th} percentile and 75\textsuperscript{th} percentiles).
    The whiskers show---according to \citet{tukey1977exploratory}---the data value just within the 1.5-fold of the inter quartile range (i.e., the difference between upper the lower quartile) exceeding the lower/upper quartiles.
    The outliers show differences in the SRTs that are larger/smaller than the ends of the whiskers.
    Top: 120 train sentences, bottom: 240 train sentences per SNR.
    }
 \label{fig:standard-boxplots}
\end{figure*}

The difference in SRT between \fastFADE{} and FADE as well as between \fastFADE{} and two empirical SRTs (from \citet{hochmuth2015influence} are depicted as boxplots in Figure \ref{fig:standard-boxplots}.
In general, SRTs simulated with \fastFADE{} were higher (worse) than with FADE.
With 120 training sentence, median differences of less than 2\,dB were found when no masker or stationary maskers were used.
The median differences were about 3\,dB with multitalker maskers and about 5\,dB, with fluctuating maskers.
However, the median differences were about 1.5\,dB when the Bisgaard profile N3 was used together with the fluctuating icra5-250m masker.
These median differences were about 1\,dB lower in all conditions when 240 train sentences were used.

In comparison with empirical measurements, \fastFADE{} predicted about 1\,dB lower SRTs with the stationary icra1m masker, and about 5\,dB higher SRTs with the fluctuating icra5-250m masker.
In silence, the SRT simulated with \fastFADE{} was about 6\,dB below the empirical SRT of 19.9\,dB\,SPL \citep{wagener2004gottinger}.
However, well-trained listeners might be able to reach such low SRTs.
For example, one subject measured by \citet{schaedler2020individual} reached an SRT in silence of about 16.5\,dB\,SPL, i.e., only 2\,dB higher than simulated with \fastFADE.
The standard deviations of the SRTs simulated with stationary (about 0.5\,dB) and fluctuating maskers (about 1.2\,dB) were on par with the inter-individual variability of the empirical data for simulations with 120 and 240 training sentences.
That is, \citet{kollmeier2015multilingual} found a standard deviation of 1.1\,dB with normal-hearing listeners in a stationary masker and \citet{hochmuth2015influence} found a standard deviation of 1.2\,dB with the icra5-250m, and of 2.9\,dB with the icra4-250m maskers.

\begin{figure}[htb]
\centering
 \includegraphics[]{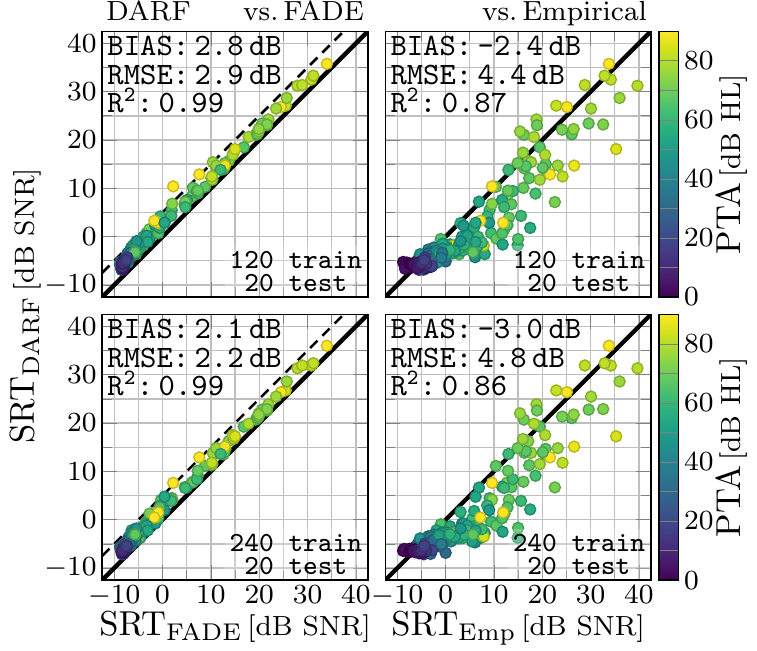}
 \caption{
  Scatter plot between SRTs simulated with \fastFADE{} (SRT$_{\text{ROR}}$), FADE (SRT$_{\text{std.}}$, left panels), and the empirical SRTs (SRT$_{\text{emp.}}$, right panels) for the data of \citet{huelsmeier2020chat}.
  The pure tone average hearing loss (PTA) was calculated across 0.5, 1, 2, and 4\,kHz.
  Note, that some of the listeners with high PTAs had steeply sloping high frequency hearing loss.
  The dashed line indicates a 5\,dB offset from the diagonal.
  Top: 120 train sentences, bottom: 240 train sentences per SNR.
 }
 \label{fig:chat-models-empirical}
\end{figure}

Figure \ref{fig:chat-models-empirical} depicts the differences in SRT between \fastFADE{} and FADE for 315 hearing-impaired ears simulated in the stationary test-specific masker (tsn) and the empirical measurements.
The simulations of both FADE versions correlate with an R$^{2}$ of 0.99.
However, the bias and RMSE indicate that hearing impairment might introduce an additional offset of about 1\,dB from standard FADE:
Both quantities exceeded the median offsets depicted in Fig. \ref{fig:standard-boxplots} by about 1\,dB.
The measured SRTs were better predicted with 120 training sentences than with 240 training sentences.
Yet, both version of \fastFADE{} showed high correlations with the empirical data as well as low bias and RMSE values.
\subsubsection{Benchmark 3: Aided hearing}
\begin{figure}[htb]
\centering
 \includegraphics[]{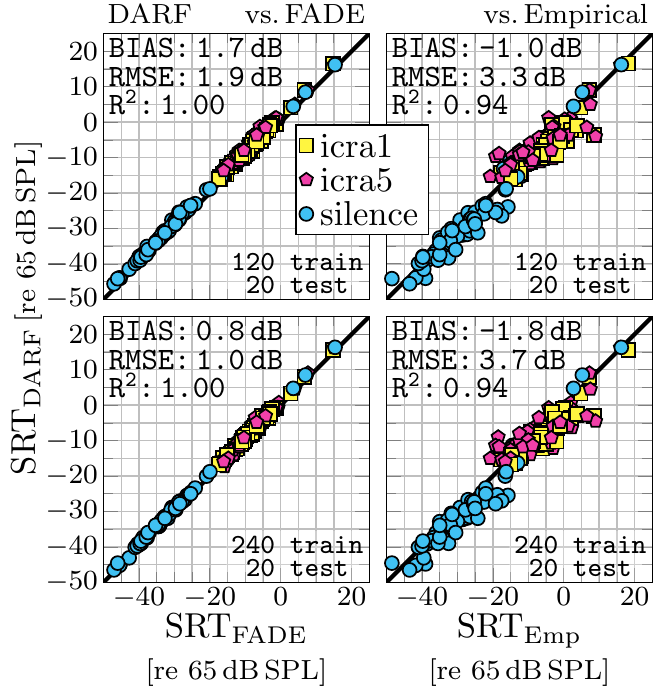}
 \caption{
  Scatter plot between SRTs simulated with \fastFADE{} (SRT$_{\text{ROR}}$), FADE (SRT$_{\text{std.}}$, left panels), and the empirical SRTs (SRT$_{\text{emp.}}$, right panels) for the data of \citet{schaedler2020individual}.
  Top: 120 train sentences, bottom: 240 train sentences per SNR, left: comparison of \fastFADE{} and FADE, right: comparison of \fastFADE{} and empirical SRTs.
  }
 \label{fig:gappp-hist}
\end{figure}

The data of \citet{schaedler2020individual} were re-simulated with \fastFADE{} with 120 and 240 train and 20 test sentences.
The differences in SRT between \fastFADE{}, FADE and the empirical measurements are depicted in Figure \mbox{\ref{fig:gappp-hist}}.
The simulations showed perfect correlations between \fastFADE{} and FADE (R$^{2}$=1.0), as well as low bias and RMSE values ($\leq$1.9\,dB) for all hearing aid algorithms.
Similar outcomes were found in comparison to the measured SRTs, i.e., SRTs predicted with both \fastFADE{} versions showed high correlations (R$^{2}\geq$0.9), as well as low bias ($\geq$-1.8\,dB) and RMSE values ($\leq$3.7\,dB) for all hearing aid algorithms.
On average, \fastFADE{} simulated higher SRTs than FADE, while it simulated SRTs lower than measured ones.
In comparison to FADE, neither hearing aid algorithms nor hearing impairment seemed to introduce further variance to the \fastFADE{} simulations for this data set, while the empirical SRTs were accurately predicted.
\subsection{Experiment 2: Benefit with real hearing aids}
\begin{figure}[htb]
\centering
 \includegraphics[]{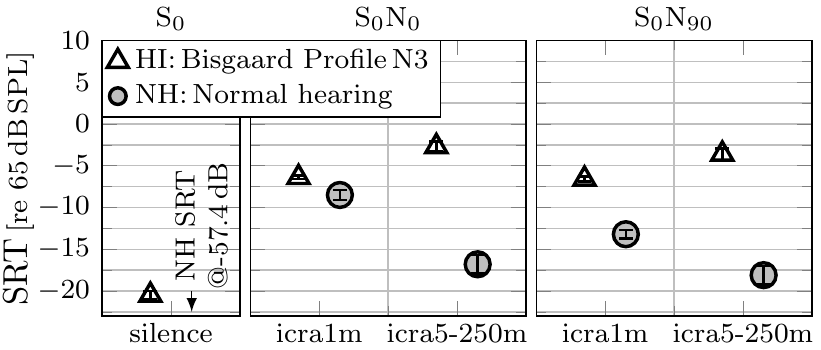}
 \caption{Simulated unaided SRTs relative to 65\,dB\,SPL in stationary (icra1m) and fluctuating (icra5-250m) maskers, and in silence with normal-hearing (NH), and assumed hearing loss \citep[profile N3 of][]{bisgaard2010standard}.
 Note that the simulated NH SRT in silence was at -57.4\,dB\,SNR relative to 65\,dB\,SPL, i.e., at about 7.6\,dB\,SPL and thus is not depicted in the panel.
 Error bars indicate the standard deviation estimated from standard deviations from the second benchmark (Fig. \ref{fig:standard-boxplots}).
 Left panel: speech from the front (S$_{0}$).
 Middle panel: speech and noise co-located from the front (S$_{0}$N$_{0}$).
 Right panel: speech from the front and noise 90$^{\circ}$ from the left (S$_{0}$N$_{90}$).
 }
 \label{fig:hearing-aid-n3-u1-srt}
\end{figure}

The SRTs simulated in the two spatial configurations without using hearing aids to process the signals are depicted in Figure \ref{fig:hearing-aid-n3-u1-srt}.
The benefits simulated with using the three hearing aid pairs are depicted in Figure \ref{fig:hearing-aid-n3-u1-benefit}.
Hearing-aid benefits were found for all devices and conditions.

\subsubsection{Unaided listening}
Generally, the unaided SRTs for the speech only S$_{0}$ and the co-located S$_{0}$N$_{0}$ conditions were at expected thresholds and matched previous simulations \citep{huelsmeier2020params} when taking the offset between \fastFADE{} and FADE into account (e.g., Fig. \ref{fig:standard-boxplots}).
The spatial separation of speech and masker (S$_{0}$N$_{90}$) led to lower SRTs when normal hearing was assumed, but seemed not to affect the SRTs simulated with hearing impairment.

As expected, the largest difference between normal and impaired hearing was found for the simulation in silence ($>$30\,dB) followed by the fluctuating masker (15\,dB) and the stationary masker (2\,dB).
These gaps increased when speech and noise were spatially separated.
The release from masking due to masker fluctuations, i.e., the difference between the stationary and the fluctuating masker, was about 9\,dB with normal hearing in the co-located, and about 5\,dB in the spatially separated condition.
This trend reversed with hearing impairment, where the benefit was about -3\,dB in both conditions.
Further, no spatial release from masking, i.e., the difference between SRTs in the co-located and the spatially separated condition, was found with the hearing-impaired configuration.
A spatial release from masking of about 5\,dB was found with the simulated normal-hearing configuration and a stationary masker, but it was only about 2\,dB with the fluctuating masker.
\subsubsection{Aided listening}
When only speech was presented from the front (i.e., speech in silence, termed S$_{0}$), the manufacturer fitting provided a benefit of about 14\,dB for all devices.
Such uniform benefits across the devices were not observed when the NAL-NL2 fitting was used.
That is, benefits of about 9\,dB were found for devices A and C, but the benefit was about 17\,dB with device B.
Nonetheless, a gap of more than 20\,dB towards simulated normal-hearing performance remained for all devices and both fitting methods.

\begin{figure}[htb]
\centering
 \includegraphics[]{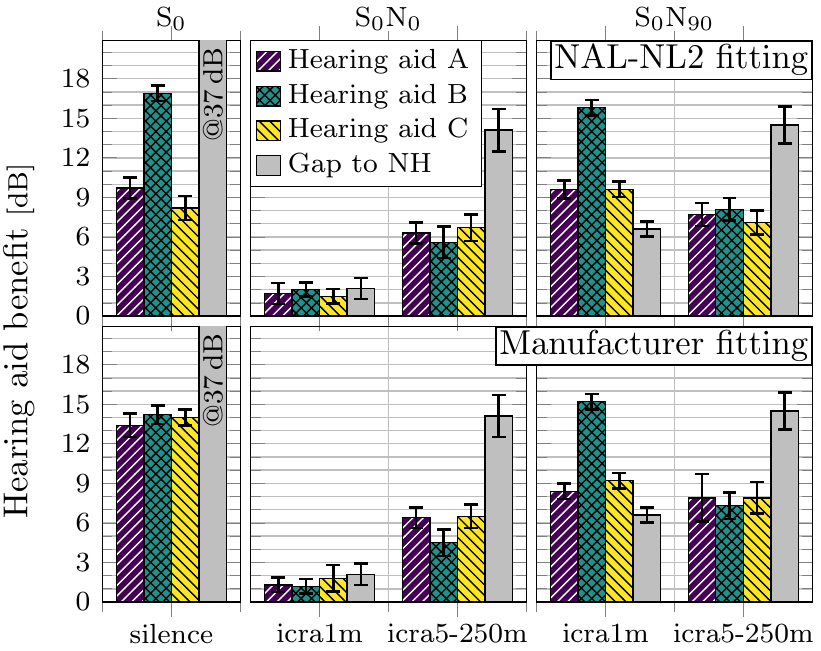}
 \caption{Simulated hearing aid benefits with NAL-NL2 fitting (second row) and the manufacturers proposed fitting method (third row).
 Simulations were conducted with assumed hearing loss \citep[profile N3 of][]{bisgaard2010standard} for three pairs of hearing aids (A, B, and C) in stationary (icra1m) and fluctuating (icra5-250m) maskers, and in silence.
 The simulated normal-hearing performance (NH) is used as fourth ``hearing aid'' (gray) to indicate the benefit required to restore normal-hearing performance (e.g., 37\,dB in silence).
 Error bars indicate the standard deviation with Gaussian error propagation for hearing aid benefits estimated from standard deviations from the second benchmark (Fig. \ref{fig:standard-boxplots}).
 Left panels: speech from the front (S$_{0}$).
 Middle panels: speech and noise co-located from the front (S$_{0}$N$_{0}$).
 Right panels: speech from the front and noise 90$^{\circ}$ from the left (S$_{0}$N$_{90}$).
 }
 \label{fig:hearing-aid-n3-u1-benefit}
\end{figure}

In the condition with co-located speech and noise from the front (S$_{0}$N$_{0}$), the benefits with the stationary icra1m masker were sufficient to restore simulated normal-hearing performance (left panels of Fig. \ref{fig:hearing-aid-n3-u1-benefit}).
A benefit between 5 and 7\,dB was simulated with the fluctuating masker, but a gap of about 7\,dB remained between the simulated aided listening performance and normal-hearing performance.
Hence, only minor differences ($\leq$2\,dB) between the fitting methods and devices were found for the co-located S$_{0}$N$_{0}$ condition.

In the condition with spatially separated speech from the front and noise from the left (S$_{0}$N$_{90}$),
benefits for speech in the stationary masker were about 10\,dB for devices A and C, but the benefit was about 16\,dB with device B for both fitting methods.
This is comparable to the simulated benefits in silence for the NAL-NL2 fitting method.
However, the aided listening performance with the stationary masker exceeded the simulated normal hearing performance at least by 1\,dB.
The simulated benefits with the fluctuating masker in the spatially separated condition were just 1 to 3\,dB better than the simulated benefits found in to co-located condition, such that a gap of about 6\,dB to normal-hearing performance remained.

\section{Discussion}
\subsection{Experiment 1: Benchmarks with digital simulation}
The first benchmark showed that \fastFADE{} can predict one SRT with less than 30 minutes of (un)processed speech signals at the cost of the simulation accuracy.
Still, \fastFADE{} even simulated 1.5\,dB higher (i.e. worse) SRTs than FADE when using 960 sentences for training and 120 for testing the speech recognizers, i.e., the configuration used with the standard version of FADE.
One reason for this discrepancy is the coarser step size of the training SNRs (\fastFADE{} 6\,dB; FADE 3\,dB), as well as the multicondition training that might not improve the recognition performance for unmatched SNR training beyond the performance already achieved when sufficient training signals are available for matched SNR training.
An optimum accuracy speed tradeoff (AST) was found for 240 train and 20 test sentences which showed the best trade-off with regard to simulation accuracy and required time.

The second benchmark showed median differences, i.e., systematic offsets, of \mbox{1 to 5\,dB} between \fastFADE{} and FADE (Fig. \ref{fig:standard-boxplots}).
Thus, a correction of these systematic offsets independent of the masker seems not to be feasible.
Predicting empirical SRTs with \fastFADE{} was accurate for the stationary icra1m masker, i.e. a systematic offset not more than 1\,dB, but not for the fluctuating masker (approx. 5\,dB offset).
There, the difference to the empirical SRT was about the same as the difference between the simulated SRTs and FADE, which implies that more training data would reduce the differences.
The standard deviations of the simulated SRTs for 120 and 240 training sentences (below 1.2\,dB for each maskers) were close to or below empirical test-retest accuracies \citep{kollmeier2015multilingual, hochmuth2015influence}.
The confidence intervals (Fig. \ref{fig:standard-boxplots}) were symmetric to the median for each masker such that random errors cancel out when sufficient repetitions are simulated.
Therefore it seems that 120 training sentences are sufficient for simulations with maskers that have little envelope fluctuations (i.e., silence or stationary maskers), or when the absolute threshold interferes with the masker's audibility (Fig. \ref{fig:standard-boxplots} and \ref{fig:chat-models-empirical}).

The simulations with \fastFADE{} of the unaided hearing impaired simulations (Fig. \ref{fig:chat-models-empirical}) showed little differences to the FADE simulations and could also predict the empirical SRTs with a high accuracy (R$^{2}\geq$0.86, RMSE$<$5\,dB).
Interestingly, \fastFADE{} provided more accurate predictions using 120 training sentences than with 240 training sentences (or than the standard FADE approach).
This, however, is only an effect of the number of used training sentences since FADE typically predicts SRTs lower than the measured SRTs.
In addition, using fewer training sentences results in an increase of the predicted SRTs (Fig. \ref{fig:map-i5-ex} and \ref{fig:standard-boxplots}).

In the third benchmark \fastFADE{} was used to simulate hearing impairment together with hearing aid algorithms (Fig. \ref{fig:gappp-hist}) that were previously recorded by \citet{schaedler2020individual}.
These simulations did not introduce further offsets than observed in the previous benchmarks.
That is, the bias across all maskers (silence, stationary and fluctuating) and hearing aid algorithms (linear amplification, compression amplification, and compression amplification with a noise-suppressing beamformer) resembled the median differences found with stationary maskers (see Fig. \ref{fig:standard-boxplots} and \ref{fig:chat-models-empirical}).
Even though fluctuating maskers were used for these simulations, the differences were not as large as found between \fastFADE{} and FADE for the second benchmark.
This might be due to the interplay between the fluctuating masker and hearing impairment, or due to the interplay between hearing aid algorithms and hearing impairment, or both.
Probably, the hearing aid algorithms also interfere with the fluctuations of the maskers:
Inaudible parts of the mixed speech and noise signal are amplified and compressed.
Consequently, the mixed signals become more spectro-temporally flat by removing spectro-temporal gaps (similar to signal representations with elevated absolute thresholds).
This simplified the training of the automatic speech recognition system since less statistic fluctuations needed to be covered by the GMMs.
Note, though, that this also prevents using the spectro-temporal gaps to recognize speech.
The comparison with the empirical measurements showed that \fastFADE{} predicted on average lower than measured SRTs, but with a high accuracy (R$^{2}\geq$0.9, RMSE$<$5\,dB).

All in all, the benchmark experiments showed that \fastFADE{} simulates similar SRTs as FADE, albeit there were larger differences between the FADE versions when a fluctuating masker was used.
Additionally, \fastFADE{} could predict the empirical outcomes with nearly the same accuracy as found with FADE, but with the exception of  fluctuating maskers with assumed normal hearing.
The third benchmark that was based on the data from \citet{schaedler2020individual} showed that the differences between the FADE versions were smaller for simulations with hearing impairment and signal processing in contrast to simulations with normal hearing and unprocessed signals.
Therefore, it seems to be plausible that \fastFADE{} can provide accurate predictions of aided speech recognition performance, especially when taking into account that \citet{schaedler2020individual} used FADE to accurately predict SRTs (R$^{2}$ = 0.94, RMSE = 4.2\,dB) and benefits (RMSE = 2.7\,dB; R$^{2}$ = 0.82) for a diverse group of listeners with and without hearing impairment.
\subsection{Experiment 2: Benefit with real hearing aids}
While a comparison between the unmodified FADE and \fastFADE{} is possible for the benchmark conditions discussed so far, this is not possible for the black-box hearing aid benefit predictions performed in Experiment 2 since not enough recorded speech material processed with the respective hearing aid algorithm was available to run the unmodified FADE.
Also, no appropriate empirical data was available to directly compare the \fastFADE{} predictions with SRT-data for individual hearing impaired subjects for the unaided and aided case.
Hence, the expected accuracy of the  predictions can only be assessed indirectly based on the following facts:
\begin{itemize}
  \item[a)]
    The expected deviation between \fastFADE{} and the unmodified FADE is small (i.e., less than 5\,dB for fluctuating maskers and less than 2\,dB for stationary maskers, see Fig. \ref{fig:standard-boxplots}), and the difference between \fastFADE{} and empirical data was small for listeners with normal hearing (5\,dB, see Fig. \ref{fig:standard-boxplots}), and also for listeners with hearing impairment (RMSE less than 5\,dB, see Fig. \ref{fig:chat-models-empirical}).
  \item[b)]
    The accuracy of the unmodified FADE for predicting individual hearing aid benefit is very high as being inferred from \citet{schaedler2020individual} in conditions with a highly controllable hearing aid \citep[MHA by][]{grimm2006mha, herzke2017open}.
\end{itemize}
Assuming that the MHA processing is comparable to the processing of the actual black-box hearing aids employed here, there is good evidence that \fastFADE{} is able to perform a precise individual patient performance prediction in accordance with the general aim of the current paper.
Moreover, the simulations performed for Experiment 2 can be assessed by considering the plausibility of the simulations in light of the current literature as follows:

The data of \citet{schaedler2020individual} included two listeners with hearing loss close to the Bisgaard profile N3.
The benefits found for one of these listeners in a spatially separated condition was 17\,dB in silence, 8\,dB in the stationary (icra1m) masker, and 8\,dB in the fluctuating (icra5-250m) maskers \citep[see the ADM \& compressive condition][Fig. 6]{schaedler2020individual}.
In Figure \ref{fig:hearing-aid-n3-u1-benefit}, the largest simulated benefits with hearing aids A and C were 17\,dB in silence, 10\,dB in stationary masker (icra1m), and 8\,dB with the fluctuating masker (icra5-250m).
Thus, the simulations with hearing aids A and C differed only by up to 2\,dB in each condition from the empirical data.
Other studies contribute to \fastFADE{}'s plausibility by reporting on equivalent aided SRTs in stationary and fluctuating maskers when speech and masker were co-located \citep[][here aided SRT stationary: -8.0\,dB\,SNR; fluctuating -8.7\,dB\,SNR]{herzke2012new}, benefits by spatially separating speech and noise when using hearing aids \citep{neher2009benefit, ahlstrom2009spatial}, or benefits found in fluctuating maskers \citep[][6 to 7\,dB]{luts2010multicenter}.

Apart from these studies, other implications for the plausibility of the simulated aided listening conditions exist.
For example, the benefits in the co-located condition were higher with the fluctuating masker than with the stationary masker.
This is plausible, since fluctuations allow for glimpsing \citep{cooke2006glimpsing}, which can be facilitated by linear amplification.
Then again, an inverse trend was observed when speech and noise were separated, i.e., the benefits with the stationary masker were higher.
This is also plausible, due to the speech-like modulations of the fluctuating masker that interfere with the beamformer and/or noise suppression algorithms.
Similar trends were observed for all three hearing aids and for both fittings.
Therefore, it is also likely to simulate similar SRTs and benefits when repeating the study.

Taken together, the simulations with \fastFADE{} seem to yield plausible outcomes, especially when taking into account the studies and benchmark experiments.
These simulations together with the accuracy of the benefits predicted by \citet{schaedler2020individual} show the feasibility of \fastFADE{} to predict the aided performance of individual listeners with different types of real hearing devices.
Even more, the simulations could demonstrate the performance differences across three types of commercial hearing devices as well.
Yet, the devices used in this study were on a comparable technological level while \citet{schaedler2020individual} only used hearing aid algorithms.
Therefore, the simulation outcome might result in less plausible outcomes for other hearing aids.
Hence, further studies need to be performed to test if the plausibility of \fastFADE{}'s predictions holds for other hearing devices as well.
\subsection{Limitations}
The real-time-optimized and data-reduced \fastFADE{} facilitates to use \textit{real} devices to process signals when simulating speech recognition thresholds without the need of empirical reference SRTs.
Therefore, the real acoustic properties and pathways of the hearing aids and the receiving dummy head can be used for the simulations.
Nonetheless, FADE is not a human recognizing speech but a machine that uses an artificial auditory system with basic binaural features.
Therefore, such simulations should always be interpreted with care.
To simulate one SRT \fastFADE{} uses less than one hour of (un)processed speech signals at the cost of a lower simulation accuracy in comparison to standard FADE.

The amount of (un)processed signals required by \fastFADE{} seems still high compared to established prediction models that typically use less than a minute of signals.
However, such models are often limited to specific speech recognition problems or require specific inputs, like, e.g., speech in stationary noise or separated/separable speech and noise signals \citep[][]{SII, beutelmann2010revision, STOI, HASPI}, which restricts their general applicability.
In order to predict aided speech recognition performance in realistic environment, models need to take into account hearing impairment, binaural hearing, realistic maskers, reverberation, and processed signals (with real devices).
For example, the SII \citep{SII} takes into account hearing impairment but neither binaural hearing nor processed signals.
BSIM \citep{beutelmann2010revision} takes into account binaural hearing but not processed signals.
HASPI \citep{HASPI} was designed to simulate aided speech recognition performance, but without taking into account fluctuating maskers \citep[see][Tab. 1 for a comparative list]{schaedler2018objective}.

Most of these model characteristics are covered with FADE and \fastFADE{} for the simulation of \textit{real} hearing aids.
However, the current version of FADE (and \fastFADE{}) neglects several binaural effects, e.g., temporal fine structure \citep{moore2012effects, neher2012binaural} or the medial olivocochlear reflex \citep{lopez2016binaural}.
Albeit it uses an automatic form of better ear listening \citep{hauth2018modeling}.
Therefore, simulations with co-located speech and noise likely yield the same outcome as monaural simulations.
Further, FADE is insensitive to the signal's overall presentation level due to the mean and variance normalization of the features, i.e., the normalization of the intensity of the signal to enhance the ASR-system's recognition performance across different SNRs, which limits its applicability.
Reverberation is another crucial component relevant in realistic scenes.
However, FADE is currently insensitive to reverberation, which is likely caused by the training and recognition method of FADE.
That is, both training and testing signals are modified with the same impulse responses such that FADE simply learns the reverberated signal representation.
Although FADE's implementation of reverberation is quite similar to that of the maskers and loss of audibility, reverberation smears the spectro-temporal energy across time in contrast to the maskers or loss of audibility which mask speech or remove energy from the signal.
Therefore, FADE's speech recognition performance is less affected by reverberation than by the maskers or loss of audibility.
A possible solution to this problem might be to use phoneme instead of word recognition.
That is, the reverberated features of a single word can be learned by the ASR system.
Presumably, this cannot be exploited when using phoneme recognition since phonemes from different words can be used for training.
That is, such phonemes have a more diverse set of preceding words that leak into the phonemes' spectro-temporal representation.
This presumably impedes the ASR system to learn the reverberated representation of single words.
\subsection{General applicability of \fastFADE{}}
In the present study \fastFADE{} was used to simulate the hearing aid benefit of three devices with similar properties.
Hence, the question arises if the approach is applicable to \textit{any} device which might process sounds vastly  different than the three examined hearing aids.
The research of \citet{schaedler2020individual} shows that FADE accurately predicts hearing aid algorithm  benefits (RMSEs below 3\,dB) which only use amplification, compression amplification, or compression amplification in combination with a noise suppressing beamformer.
Most hearing aid fitting rules apply a combination of amplification and compression \citep{valente1998guidelines, keidser2011nal, oetting2018restoring}.
Hence, \fastFADE{} might provide accurate hearing aid benefit predictions for devices which amplify and compress sound.

Yet, hearing aids do more than just amplification and compression.
That is, hearing aids typically support different modes (e.g., directional vs. omnidirectional amplification) that can be selected by the hearing aid user, or that are selected by the hearing aids dependent on the acoustic environment \citep{krystek2016smart}.
Benefit predictions for devices that support such modes and which are used in more realistic environments likely differ from the \fastFADE{} simulations made in a soundproof listening booth, even though the same hearing aids are used.
\subsection{Applications of \fastFADE{}}
FADE was successfully modified to reduce simulation time in conditions where only real time signal processing is available.
Yet, the benchmarks showed that random and static offsets between the FADE versions remain, especially when fluctuating maskers are used.
This implies that the simulation accuracy might decrease when the acoustic scene becomes more complex.
In such cases, a more conservative approach with \fastFADE{} might be required.
For example, more training sentences and a finer grid of training SNRs could be used when complexity increases.
In such cases, a preliminary fine tuning of \fastFADE{} should be conducted, e.g., by determining a required accuracy, or by selecting optimal ASTs for each parameter and masker.
Nevertheless, it is assumed that the parameter space exploited in the current paper should provide a sufficient guideline for most applications.

Note that an accuracy-time trade-off exists for the ASR-based approaches that predict aided performance.
While the approach by \citet{fontan2020improving} provides approximate recognition rates for a given condition thus requiring little execution time while being susceptible to ceiling effects due to using fixed SNRs \citep[see also][]{fontan2020predicting}, the complete FADE model avoids this by estimating the SRT using several recognition rates at different SNR conditions which requires much more processing time.
Both approaches facilitate faster simulations than \fastFADE{} when the signals can be processed offline, i.e., without the constraint that signals have to be recorded.
However, this requires modeling the acoustic path of the signals as well as other physical properties of the hearing aids which is avoided by the \fastFADE{} approach.

Possible fields of application for \fastFADE{} include hearing aid development, the selection of individual hearing aids, or the optimization of hearing aid fittings with respect to speech recognition performance.
\fastFADE{} can provide guidance for the development of hearing aids by indicating which algorithm provides benefits and which one does not.
This might even be used in more complex, but realistic and relevant acoustic environments, such as, e.g., in cafeterias, bars, or supermarkets.
Similarly, the process of finding an optimal hearing aid for an individual \citep[which may take some weeks, see][]{boymans2012audiologist} might be accelerated by using \fastFADE{} or other objective pre-selection methods of hearing aids.
That is, such models could be individualized with the listener's hearing impairment to perform simulations with different hearing aids to find the optimum device and parameter setting.
Models---unlike listeners---neither fatigue nor do they need to acclimatize to the hearing aids which allows to make several simulations without needing the patient to be present.
This might reduce the required number of iterative visits to audiologists while increasing the acceptance of the devices simply by improving the hearing aid recommendation procedure.
Further, this approach could be extended to provide a first fit of the hearing aids as proposed by \citet{volker2018hearing}.
Since FADE currently does not take into account overall presentation levels, or rather loudness, its applicability to automatic hearing aid fitting is limited.
Yet, \fastFADE{} could be paired with a loudness model \citep[e.g.,][]{moore2004revised, moore2007modeling, oetting2016spectral, pieper2018physiologically} to predict if a proposed fitting method is acceptable.

An additional benefit of using \fastFADE{} is its direct applicability to all languages for which a matrix sentence test was developed (i.e., more than 20).
This facilitates to objectively compare \textit{different} devices used in \textit{different} maskers and \textit{different} languages.
Therefore, differences in required properties of hearing aids in dependence on the language would become apparent as long as they relate to those acoustic properties predictable by FADE.
For example, the benefit provided by hearing aids with the same setting for tonal languages like Mandarin or Cantonese might differ significantly from benefits found for non-tonal languages like English or German.
Note, however, that FADE appears to utilize the respective optimum speech features in each language since the deviation between simulations and empirical data is the same in a tonal language (Mandarin: empirical -11.2 dB \citep{hu2018construction}, FADE: -12.7\,dB, data not published) as in German, Polish, Russian, and Spanish SRTs \citep{schaedler2016microscopic}.
\subsection{Comparison with ASR-based models from the literature}
Currently, only few models use ASR to predict the (aided) speech recognition performance of hearing impaired listeners, these are:
a) \citet{fontan2017automatic} uses a GMM-HMM ASR system trained with several hours of noisy speech signals to predict a number of correctly recognized words for a fixed SNR.
b) \citet{spille2017predicting} uses a deep neural net HMM ASR system trained with several hours of noisy speech signals to predict SRTs obtained with sentences.
c) FADE (described here) uses GMM-HMM ASR systems where each is trained with about 45 minutes of noisy matrix sentences to predict SRTs \citep{schaedler2015matrix}.

The model of \citet{fontan2017automatic} \citep[see also][]{fontan2020improving, fontan2020predicting} predicts speech recognition scores of listeners with (simulated) hearing impairment.
Since a non-specialized ASR algorithm was employed, a typical human-machine gap is observed \citep{spille2018comparing}, i.e. the ASR-based model predicts lower percentages of correctly recognized words than measured with the human listeners.
Yet, the predictions showed high Pearson \citep[][]{fontan2017automatic} or Spearman correlations \citep[][]{fontan2020predicting} with the human data.
Later, \citet{fontan2020improving} used the model to predict ``optimized'' hearing aid gains generated for 24 listeners with age-related hearing loss.
These hearing aid gains led to 1.0 to 2.4 percent points (pp) higher speech recognition performance scores than with the CAM2 prescription rule \citep{moore2010development}.
However, such small differences are most likely not significant on an individual level and may therefore exhibit little utility for hearing aid fitting or predicting hearing aid benefit to an indivudal listener.

Rather, these small differences are a typical problem of floor or ceiling effects that may occur if measuring or modeling recognition rates for fixed SNRs.
Specifically, both prescription rules of \citet{fontan2020improving} resulted in scores close to 100\,\%.
To avoid such floor and ceiling effects and to achieve a better discrimination across different hearing aid settings, adaptive tracking procedures are commonly used \citep[e.g.,][]{levitt1971transformed, plomp1979improving} that converge on a fixed percentage on the discrimiation function which should be as steep as possible \citep[e.g.,][]{kollmeier2015multilingual}.
Hence, it might be useful in future work to modify the approach by Fontan et al. (2020a) for predicting the SRT either by interpolating between the recognition performances for several SNRs or by directly combining it with an adaptive tracking procedure.
This might facilitate the direct comparison of its respective performance  with other ASR-based models.

\citet{spille2017predicting} predicted SRTs of listeners with normal hearing with a high accuracy.
That is, the absolute deviation between human performance and prediction was less than 2\,dB on the average \citep[see][Tab. 5, col \textit{Multi}]{spille2017predicting}.
This nearly vanishing human-machine gap was possible due to the training procedure and the limited target sentence materials: For training, ten hours of German matrix sentences recorded with 20 speakers were mixed with eight different maskers (in total 80) and for testing each masker was mixed with 3200 sentences of the original German matrix sentence test at different SNRs.
Later, \citet{spille2018comparing} predicted SRTs of hearing impaired listeners, but mainly to examine the influence of the acoustic complexity on the recognition rate for human listeners and their ASR system.
However, this model was not evaluated for predicting hearing aid algorithm benefits.

Previous research with FADE included the simulation of SRTs of listeners with normal \citep[predictions within empirical SRT range][]{schaedler2016simulation} and impaired hearing \citep{schaedler2018objective}, and the prediction of hearing aid algorithm benefits \citep[RMSEs$\leq$3.5\,dB][]{schaedler2018objective, schaedler2020individual}.
FADE completely eliminates the human-machine gap by using the same speech signals (but different noise tokens) for training and testing the ASR system which is supposed to represent a listener who is familiar with the speech material.
This requires to re-train the model for each SRT simulation.
However, training the ASR systems of FADE is by far less computational intensive than for the systems of \citet{fontan2017automatic} or \citet{spille2017predicting}, which allows the application of FADE to matrix sentence tests in different languages without changing the model structure.
Nonetheless---and as shown by \citet{schaedler2020individual}---this results in very accurate SRT and hearing aid algorithm benefit predictions for individual listeners.

Taken together, a direct model comparison of the model employed here with other models from the literature is difficult because different prerequisites and evaluation schemes have been employed.
Nevertheless, it is fair to say that FADE predicts (aided) SRTs for listeners with and without hearing impairment, while DARF facilitates SRT benefit predictions with real hearing aids in a reasonable time frame.

\section{Conclusions}
The most important findings and implications of this study are:
\begin{enumerate}
  \item
    The data-reduced version of FADE (\fastFADE{}) facilitates simulations with a reduced number of train and test signals in contrast to the original FADE.
    \fastFADE{} can be used with real hearing devices with unknown properties (i.e., signal processing black boxes) to simulate one SRT with about 30 minutes of mixed signals of matrix sentence tests.
    Note that a validation of this approach with comparison between prediction and actual individual performance is still open.
    Hence, a thorough examination of the simulation outcome is required.
    Nonetheless, \fastFADE{} enables in-situ objective evaluations of hearing aids or prototypes thereof.
  \item
    The deviation from predictions with the unmodified FADE approach in the proposed configuration (120 or 240 train and 20 test sentences per SNR) was found to be small (1 to 2\,dB) in many cases, and as large as 4 to 5\,dB in a strongly fluctuating noise masker.
    The flexibility of the approach allows to reduce the deviation (to 1.5\,dB) by increasing the simulation time (to about 2.5 hours with 960 train sentences per SNR).
  \item
    The modified approach enables an individual, nonintrusive hearing aid benefit prediction based on a limited amount of recorded signals using the respective hearing aid  as ``black box'' device.
    The accuracy of the unmodified FADE for predicting hearing aid benefits \citep{schaedler2020individual} and the plausibility of the prediction demonstrated here hint towards the potential of this approach for model-based hearing aid fitting.
    Likewise, the development of hearing devices might benefit from \fastFADE{}.
\end{enumerate}

\section*{Funding}
This work was supported by the Cluster of Excellence Grant ``Hearing4all'' (DFG Project Number 390895286).
\section*{Acknowledgments}
We thank Florian Denk of the University of Oldenburg for his kind help with setting up the dummy head, the H\"orzentrum Oldenburg for providing the hearing aids, and especially Kevin Pollak of the H\"orzentrum Oldenburg for fitting the hearing aids.

\section*{Declaration of Conflicting Interests}
The Authors declare that there is no conflict of interest.

\bibliography{sources-relevant}
\cleardoublepage
\section{Supplemental Material}
\begin{figure}[htb]
\centering
 \includegraphics[]{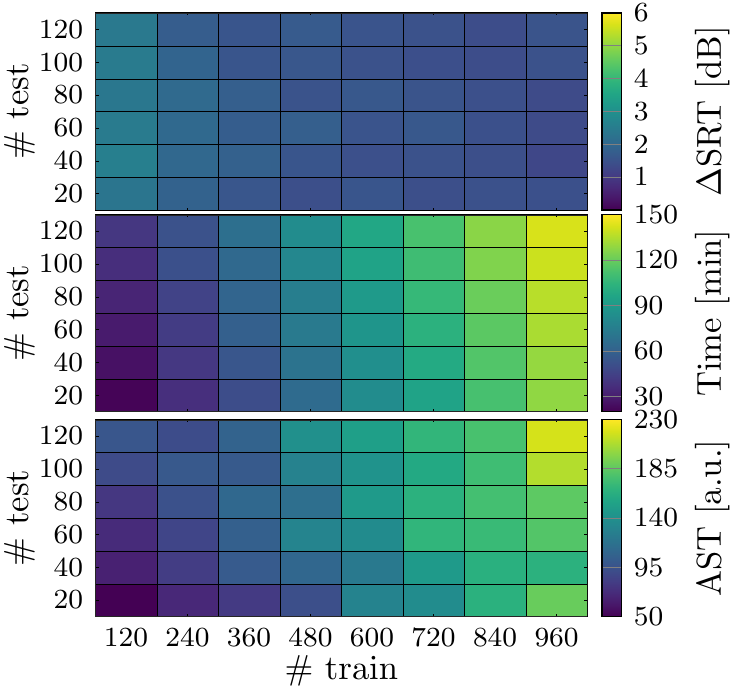}
 \caption{%
    Model performance comparison for the stationary icra1m masker for different numbers (\#) of train and test sentences (average of 16 simulations).
    Top: Differences in SRT between \fastFADE{} and FADE.
    Middle: Accumulated duration of all used sentences.
    Bottom: Accuracy speed tradeoff measure (AST).
    }
 \label{fig:map-i1}
\end{figure}

\begin{figure}[htb]
\centering
 \includegraphics[]{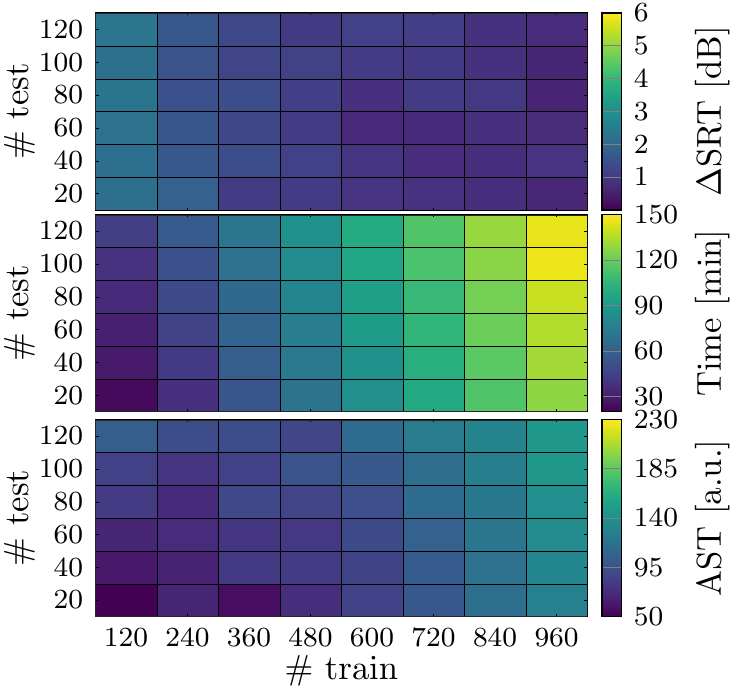}
 \caption{%
    Like Figure \ref{fig:map-i1}, but for a 20-talker babble noise masker.
    }
 \label{fig:map-bab}
\end{figure}

\begin{figure}[htb]
\centering
 \includegraphics[]{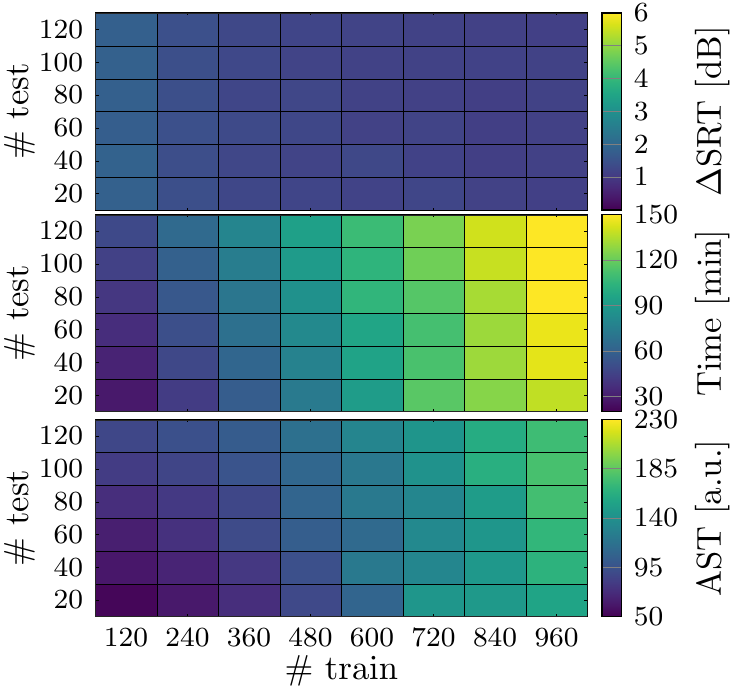}
 \caption{%
    Like Figure \ref{fig:map-i1}, but without masker (silence).
    }
 \label{fig:map-sil}
\end{figure}

\begin{figure}[htb]
\centering
 \includegraphics[]{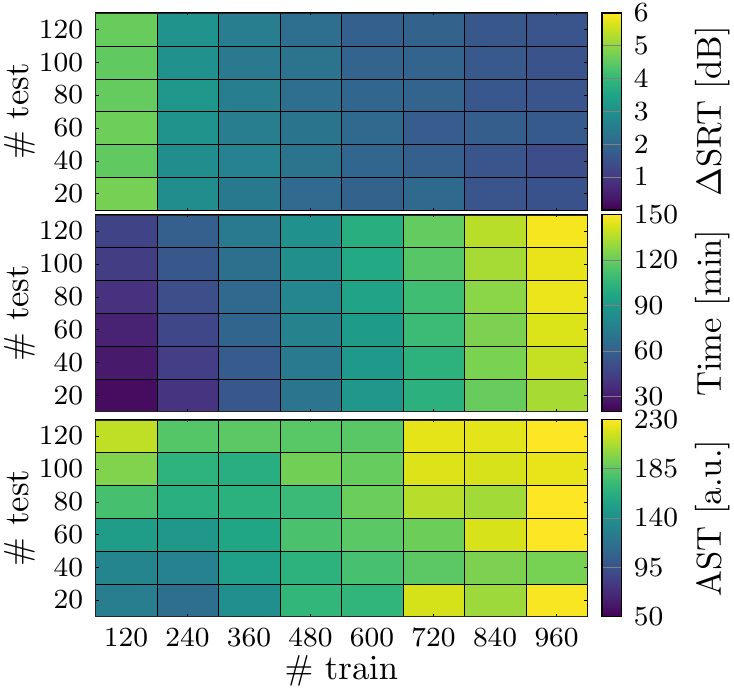}
 \caption{%
    Like Figure \ref{fig:map-i1}, but for the icra5-250m masker and based on 64 simulations.
    Note that the AST data depicted here has the same scaling as Figures \ref{fig:map-i1} to \ref{fig:map-sil}.
    }
 \label{fig:map-i5}
\end{figure}

Figures \ref{fig:map-i1}, \ref{fig:map-bab}, and \ref{fig:map-sil} show the same simulations as conducted for the fist benchmark experiment but for the stationary icra1m masker (Fig. \ref{fig:map-i1}), the 20 talker babble noise masker (Fig. \ref{fig:map-bab}), and silence (Fig. \ref{fig:map-sil}).
Figure \ref{fig:map-i5} shows the data for the icra5-250m masker as used in the first benchmark, but with the same range of ASTs as in the other figures shown here.


\end{document}